\documentclass
[superscriptaddress,secnumarabic,amssymb,amsmath,nobibnotes,aps,prd,showkeys,showpacs,twocolumn,nofootinbib]{revtex4}%
\usepackage{bm}
\usepackage{hyperref}
\usepackage{mathrsfs}
\usepackage{xcolor,color,graphicx,graphics}
\usepackage[all]{xy}
\usepackage{latexsym,amssymb,amsmath,amsfonts} 
\usepackage[english]{babel} 
\usepackage[OT1]{fontenc}
\usepackage[latin1]{inputenc}
\usepackage{makeidx}
\usepackage{hyperref}
\usepackage{color,graphicx,graphics,wrapfig,epsf}

\usepackage{subfig}
\usepackage{mwe}
\hypersetup{urlcolor=BlueViolet,
	    citecolor=Plum,
	    linkcolor=PineGreen}

\begin{document}

\title{
Earthquakes as probing tools for gravity theories}

\author{Aleksander Kozak}
\email{aleksander.kozak@uwr.edu.pl}
\affiliation{Institute of Theoretical physics, University of Wroclaw, pl. Maxa Borna 9, 50-206 Wroclaw, Poland
}

\author{Aneta Wojnar}
\thanks{Corresponding author}
\email[E-mail: ]{awojnar@ucm.es}
\affiliation{Departamento de F\'isica Te\'orica \& IPARCOS, Universidad Complutense de Madrid, E-28040, 
Madrid, Spain}

\begin{abstract}
We propose a novel method for testing gravity models using seismic data from Earth. By imposing observational constraints on Earth's moment of inertia and mass, we rigorously limit the gravitational models' parameters within a $2\sigma$ accuracy. Our method constrains the parameters governing additional terms to the General Relativity Lagrangian to the following ranges: $-2\times10^9\lesssim\beta\lesssim 10^9 \text{m}^2$ for Palatini $f(R)$ gravity, $-8\times10^9\lesssim\epsilon\lesssim 4\times 10^9 \text{m}^2$ for Eddington-inspired Born-Infeld gravity, and $-10^{-3}\lesssim\Upsilon\lesssim10^{-3}$ for Degenerate Higher-Order Scalar-Tensor theories. We also discuss potential avenues to enhance the proposed method, aiming to impose even tighter constraints on gravity models. 
\end{abstract}

\maketitle

Several proposals have been made to extend General Relativity (GR) for addressing the mysteries of the dark sector in the Universe \cite{huterer,cantata}. These modifications are necessary at the cosmological level but should be suppressed in small-scale systems like compact objects and the Solar System. Some of the most popular modifications of GR, Degenerate Higher-Order Scalar Tensor (DHOST) theories \cite{langlois} and Ricci-based theories \cite{ delhom2019}, circumvent the problem of the impact of modifications of gravity on small-scale objects by either hiding the effect of an additional degree of freedom via screening mechanisms, or reduce to GR with the cosmological constant in the vacuum. DHOST theories, featuring a dynamical scalar field, utilize the Vainshtein mechanism \cite{ babichev2013}, so that the interactions with the field become noticeable only at the cosmological scale and inside astrophysical bodies \cite{koy}. Ricci-based theories, on the other hand, do not introduce any additional propagating degrees of freedom. The modifications turn out to depend on the trace of the energy-momentum tensor of  baryonic matter, therefore the theories simplify to GR with the inclusion of the cosmological constant in the empty or radiation-dominated spacetime. 

A recent analysis of the gravitational parameter space \cite{baker} has unveiled unexplored regions that correspond to galaxies and stellar objects. These untested domains, which delineate small-scale systems from the cosmological regime, offer the potential for gaining insights into corrections to GR. To address the existing gap and perform comprehensive tests of gravitational proposals, we introduced a new approach based on planetary seismology \cite{olek1,Kozak:2023axy}. Previously, seismic data from low-mass stars (asteroseismology) \cite{bell} and the Sun (helioseismology) \cite{saltas1, saltas2} were utilized for constraining fundamental theories. The increasing quantity and precision of observational data on astrophysical bodies have facilitated the constraints or rejection of certain gravity theories. For instance, Multi-messenger Astronomy \cite{ab1, pat} has ruled out models predicting different speeds for gravitational waves and light \cite{baker2, ez2}. Additionally, soft equations of state, which fail to support high neutron star masses within the framework of GR, have been ruled out \cite{ rez, fon, cro}, although they remain viable descriptions in modified gravity proposals \cite{review}.

In contrast to compact and stellar objects, where equations of state and atmospheric properties introduce uncertainties \cite{olek4,debora,wojnar}, Earth seismology offers insights into the planet's interior \cite{poirier,prem,kus,ken,ken2,iris}. Utilizing seismic data, along with precise measurements of Earth's mass and moment of inertia, provides a powerful tool to constrain gravity models, leveraging well-understood physics and mitigating some uncertainties associated with model assumptions.

Recent advancements in seismographic tools \cite{prem, mush, frost, step, pham} and laboratory experiments simulating extreme temperatures and pressures in Earth's interior have significantly enhanced our understanding of Earth's interior properties, particularly those of iron and its compounds \cite{laser}. Additionally, new neutrino telescopes offer information on density, composition, and abundances of light elements in the outer core, further reducing uncertainties related to Earth's core characteristics \cite{top}.

However, concerns about the negligible impact of modified gravity effects in stellar and planetary physics arise. Although the influence on layer densities and thicknesses is small, it remains significant \cite{olek2, olek3, wojnar2}. Our extensive and accurate knowledge of the Solar System planets, particularly Earth \cite{ bill, folk, konopliv, smith, folk2, ziemia}, enables us to utilize available data for constraining theories \cite{olek1,Kozak:2023axy}. In our simplified approach, we achieve accuracy up to $2\sigma$ level.

Some of the existing gravitational proposals introduce a correction term to the Poisson equation\footnote{That is, the correction term is also of the second order in velocities.}.
In what follows, let us focus on such theories of gravity whose Poisson equation can be written as 
\begin{equation}\label{poisson}
    \nabla^2\phi(\mathbf{x}) =  4\pi G\Big(\rho(\mathbf{x})+\nabla^2\alpha\big(\mathbf{x},\rho(\mathbf{x})\big)\Big),
\end{equation}
where $\alpha(\mathbf{x},\rho(\mathbf{x}))$ is in principle an arbitrary function of position $\mathbf{x}$ in given coordinates and matter fields represented by the energy density $\rho(\mathbf{x})$. We will restrict our considerations to the spherical-symmetric case such that all quantities are (radial distance) $r-$dependent. Then, the above Poisson equation includes theories  such as DHOST \cite{koy} with the correction of the form $\alpha(r,\rho)=\frac{\Upsilon}{4}r^2\rho(r)$, with $\Upsilon$ being a constant parameter, as well as Ricci-based gravity \cite{banados,toniato}. Regarding the last one, we will focus on Palatini $f(R)$ gravity, providing the correction $\alpha(r,\rho)=2\beta\rho(r)$, where $\beta=\textrm{const}$ is a parameter accompanying the quadratic term in the gravitational Lagrangian, and EiBI gravity, with the correction of the form $\alpha(r,\rho)=\epsilon/2 \rho(r)$, where $\epsilon$ is a constant theory parameter. Since the relation between parameters of Palatini and EiBI is $\epsilon=4\beta$, in the further part we will deal with only two parameters, $\Upsilon$ for DHOST and $\beta$ for Ricci-based. 

In those theories,
the non-relativistic hydrostatic equilibrium and mass equations in the spherical-symmetric case are
\begin{align}
    \frac{d\phi}{dr}&=-\rho^{-1}\frac{dP}{dr}\label{hydro},\\
        M&=\int 4\pi'\tilde r^2 \rho(\tilde r) d\tilde r\label{mass},
\end{align}
where $P$ is pressure and $M$ mass included in a ball within the radius $r=R$. 
Similarly to the 1-dimensional Preliminary Reference Earth Model (PREM) \cite{prem}, we will assume that one deals with the adiabatic compression such that no exchange of heat between the Earth's layers takes place. Apart from it, the planet is in the hydrostatic equilibrium described by \eqref{hydro}, with radially symmetric shells with the given density jump between the inner and outer core $\Delta\rho=600$, central density $\rho_c=13050$ and density at the mantle's base $\rho_m=5563$ (in kg/m$^3$). On the other hand, the densities in the outer layers are given by the empirical Birch's law $ \rho = a + b v_p$, with $a$ and $b$ being parameters which depend on the mean atomic mass of the material in the upper mantle \cite{prem}.
The longitudinal elastic wave $v_p$, together with the transverse elastic wave $v_s$ allows to define the seismic parameter $\Phi_s$ as \cite{poirier} 
    \begin{equation}\label{seismic}
        \Phi_s= v_p^2 - \frac{4}{3} v_s^2.
    \end{equation}
    \begin{figure}[t]
\centering
\includegraphics[scale=.51]{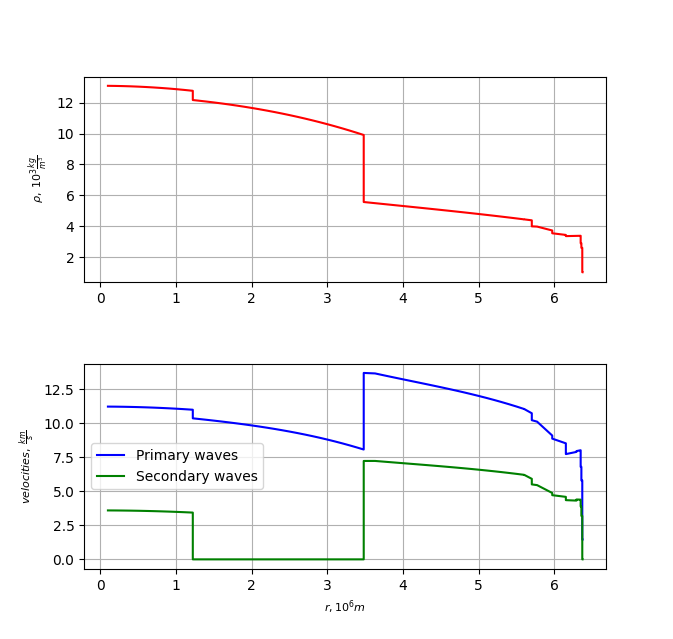}
\caption{[color online] The density profile of the Earth according to the Preliminary Reference Earth Model (PREM) \cite{prem,Kozak:2023axy}, which assumes Newtonian gravity. The velocity plots are derived from observational data without incorporating any gravitational theory. The primary waves refer to longitudinal waves, while the secondary waves are transverse waves. }
\label{fig1}
\end{figure}
 Those velocity-depth profiles, see Fig. \ref{fig1}, are given by the travel-time distance curves for seismic waves and on periods of free oscillations \cite{poirier, bolt, bullen}. Using a hydrostatic equilibrium equation, they provide pressure, density and elastic moduli profiles as functions of depth.  

On the other hand, the seismic parameter \eqref{seismic} is related to the elastic properties of an isotropic material, that is, to the bulk modulus $K$ (incompressibility) 
\begin{equation}
     \Phi_s=  \frac{K}{\rho}.
\end{equation}
Applying the definition of the bulk modulus $ K= \frac{dP}{d \mathrm{ln}\rho}$,
the seismic parameter can be written in terms of the material's properties
\begin{equation}\label{eos}
     \Phi_s=  \frac{dP}{d\rho},
\end{equation}
such that is it clear now that it includes information on the equation of state. We can then use it in \eqref{hydro} to write
  \begin{equation}\label{hydro2}
        \frac{d\rho}{dr} =  -\rho  \Phi_s^{-1} \frac{d\phi}{dr}.
    \end{equation}
Mass equation \eqref{mass} and moment of inertia 
\begin{equation}
    I= \frac{8}{3}\pi \int_0^R r^4\rho(r) dr,
\end{equation}
play a role of the constraints whose values are given by observations with a high accuracy \cite{luzum,chen}.

\begin{figure*} 
  \centering
  \advance\leftskip-0.7cm
 \subfloat{\includegraphics[width=.51\linewidth]{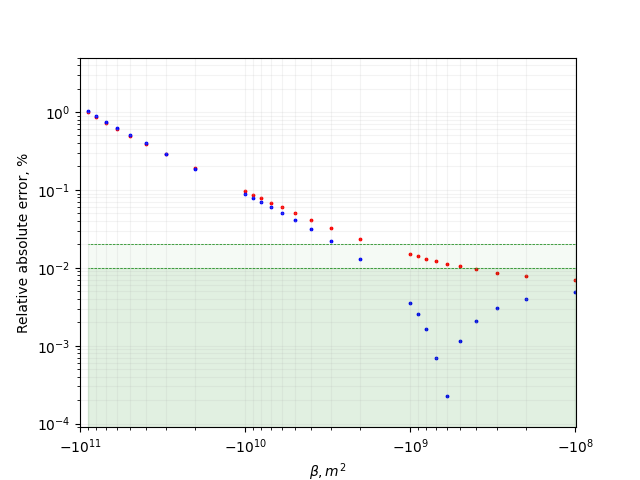}\quad\includegraphics[width=.51\linewidth]{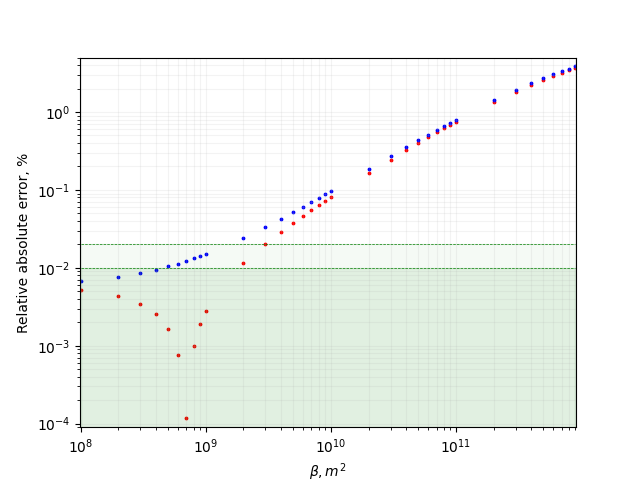}}\\[\baselineskip]
  \subfloat{\includegraphics[width=.51\linewidth]{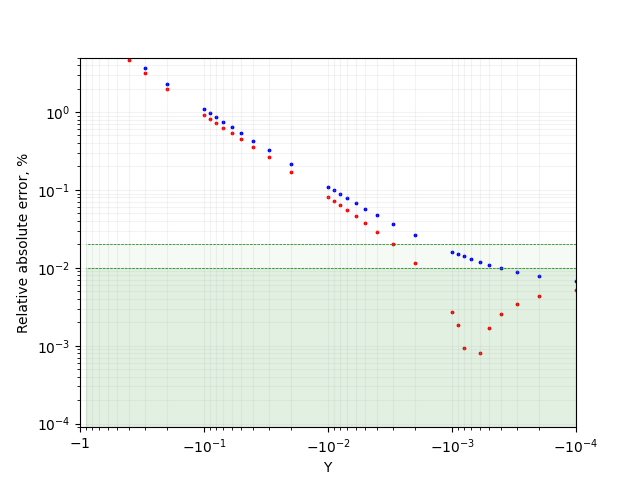}\quad\includegraphics[width=.51\linewidth]{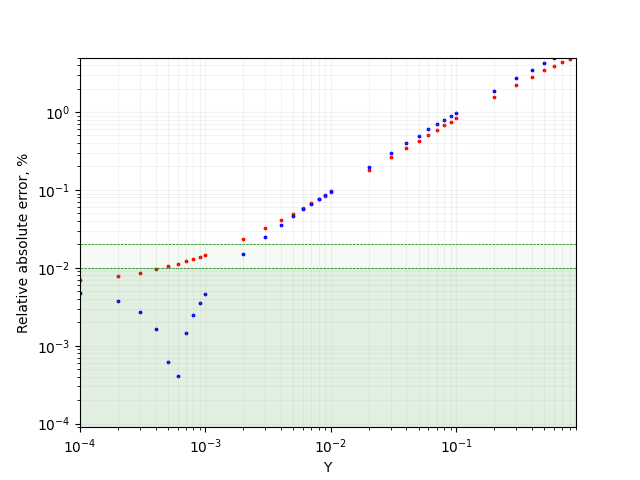}}
\caption{[color online] The relative absolute errors for Earth's mass and moment of inertia. Red dots represent moment of inertia errors, while blue dots represent mass errors. The dark green stripe represents the 1-sigma region, and the light green denotes the 2-sigma region. The green region encompasses uncertainties for both mass and moment of inertia, as their ratios of sigma to mean value are similar (approximately 0.01\%).}
\label{results}
\end{figure*}

To go further, we base our calculations on the data set provided by \cite{prem} and references therein, which includes measured values of seismic waves velocities. By assuming values for the free parameters of the PREM model, i.e. the density at the case of the mantle $\rho_m$, the density at the core $\rho_c$, and the density jump between the inner and the outer core $\Delta\rho$, we calculated the density profiles and obtained the total mass and polar moment of inertia. We aimed to obtain values that were consistent with those predicted by PREM, while accounting for the uncertainties arising from measurements. Our calculations were performed using a Python script. We varied the values of $\beta$  and $\Upsilon$ and observed the effects on the calculations. We used a simplified model to assess the crucial parameters for more sophisticated analysis and to determine the order of magnitude of $\beta$ and $\Upsilon$ at which the effects of modified gravity are still in agreement with the observed constraints. Our results are shown in Fig. \ref{results}. The integration technique involved fitting a curve to the data points when finding the relation between the depth and the seismic parameter, and using the Euler method with initial conditions at corresponding boundaries. We then computed the errors for a fixed set of parameters.

\begin{figure*} \label{1e8}
  \centering
  \advance\leftskip-1.5cm
  \subfloat{\includegraphics[scale=0.5]{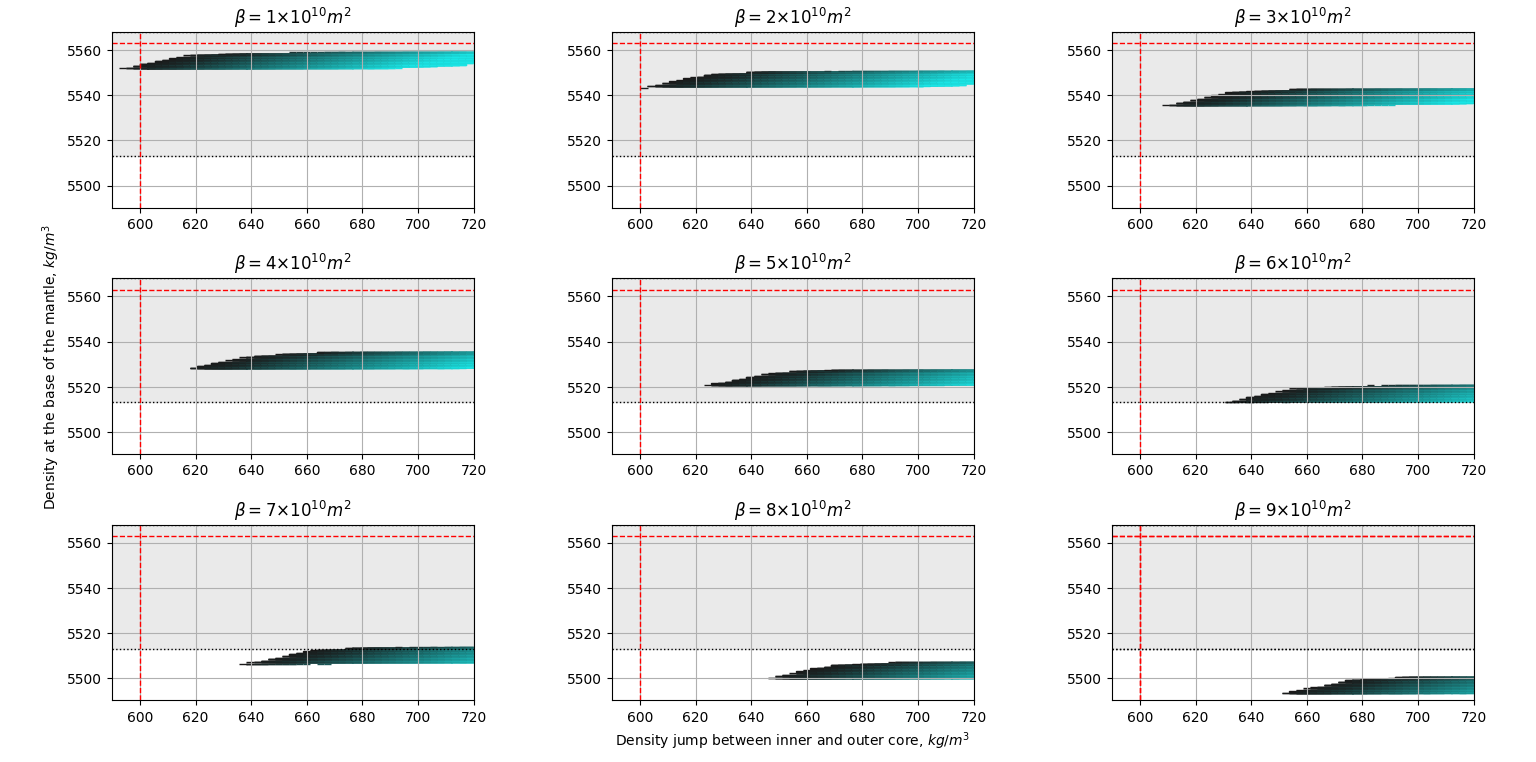}} \\
  \subfloat{\includegraphics[scale=0.7]{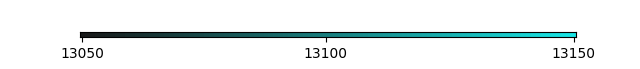}}
  \caption{[color online] $2\sigma$ confidence regions of the theory parameters $(\rho_c, \rho_m, \Delta \rho)$ for different values of the $\beta$ parameter, being of order $10^{10}$ m$^2$. Lower values of the central density are related to darker shades of blue, while the larger values - to lighter; the range of the central density values, together with the corresponding colors, is shown in the color bar below the figures. The units are kg/m$^3$. The red dashed lines represent the PREM values of the density jump and the density at the base of the mantle. The grey region corresponds to the maximum uncertainty in the determination of the density at the base of the mantle $\rho_m$, which was assumed to be $50 \text{ kg m}^{-3}$ (with respect to the PREM value). }
  \label{e10}
\end{figure*}

It must be stressed that the central density's value, unlike in the PREM model, is not a result of solving differential equations, but rather an initial assumption. Additionally, we are not interested in the outermost layers, since we expect the modifications of gravity to be weak there. For this reason, we assume Birch's law to hold, and all the density values are taken directly from the PREM model. Then, we integrated \eqref{poisson}, together with the mass relation \eqref{mass}, for different values of theories' parameters. For the Ricci based gravity, we performed calculations for $\beta \in [-10^{11}, 10^{11}]\, \text{m}^2$, while for the DHOST theories, the range of the theory's parameter was $\Upsilon \in [-1, 1]$. The measure of goodness of the model with a particular choice of the corresponding parameter was the extent to which the calculated mass and polar moment of inertia agreed with the measured ones: $M_\oplus = (5.9722 \pm 0.0006)\times 10^{24} \text{kg}$ \cite{luzum} and $I_\oplus = (8.01736\pm 0.00097)\times 10^{37} \text{kg m}^2$ \cite{chen}. 

Our study demonstrates that the following bounds can be placed on the considered parameters, ensuring that the deviations of Earth's mass and polar moment of inertia do not exceed $2\sigma$. For Palatini $f(R)$ gravity, this value is approximately  $-2\times10^9\lesssim\beta\lesssim 10^9 \text{m}^2$. Similarly, for EiBI gravity, the value is around $-8\times10^9\lesssim\epsilon\lesssim 4\times 10^9 \text{m}^2$. In the case of DHOST model, the constraint is $-10^{-3}\lesssim\Upsilon\lesssim10^{-3}$.

{
Considering PREM as a valid model for Earth, the uncertainties in the moment of inertia and mass provide bounds for the parameters. However, since PREM is not a perfect model, density parameters can differ from our assumptions. Notably, there exists a region for a given theory parameter where all three density parameters agree with experimental measurements (see our analisys in \cite{Kozak:2023axy}). Among them, $\rho_m$ has a narrower range of variation compared to $\Delta\rho$ and $\rho_c$. 
As an example, for $\beta=10^9 \text{m}^2$, the deviations in $\rho_m$ required to maintain the same mass and polar moment of inertia, compared to the case when $\beta=0$ ($\Upsilon=0$), are very small, amounting to only $0.02\%$. However, the worst-case uncertainty in the PREM model, with deviations of about $50 \,\text{kg m}^{-3}$, is $0.9\%$, while keeping $\Delta\rho$ and $\rho_c$ unchanged. For such a deviation, the parameter increases by almost two orders of magnitude, as demonstrated in Fig. \ref{e10}. This shows that the impact of changing the theory parameter on $\rho_m$ is smaller than the uncertainty in the PREM model, which implies that the effect of the parameter on the results is relatively small compared to the uncertainties in the Earth model itself. Therefore, since the impact of the gravitational parameters on the possible range of $\rho_m$ is significant, reducing the uncertainty of $\rho_m$ by using a better Earth model will further improve our constraints on gravity models. Note that in this work we constrain models of gravity using the density parameters known from PREM, assuming that it is a viable model of Earth.
}


Our approach has limitations arising from assumptions and simplifications. However, to enhance this research, several potential extensions can be explored. The primary concern is assuming spherical symmetry, which fails to consider Earth's actual shape and its sensitivity to rotation. To overcome this, we will focus on estimating the equatorial moment of inertia compared to the polar moment by incorporating travel time ellipticity corrections into the PREM model \cite{ken,ken1} in the near future.

Additionally, both PREM and our models are one-dimensional with spherical layers, which affect the moment of inertia and mass. Our future work should also consider the imperfections and density variations.

It is important to note that PREM does not consider seismic wave travel times sampling the boundaries of the outer and inner core, making it unsuitable to described the deepest part of the planet. In order to take it into account, we will use the more accurate model AK135-F, which accounts for core wave complexity \cite{kennett,mont}. Moreover, equations of state can also be used to model core density and bulk moduli \cite{irving} rather than relying solely on seismic data.

Regarding Birch's law which describes the outer layers, it is an empirical law with experimentally obtained coefficients, it can be safely used in our case. However, if dealing with seismic data from Mars \cite{irv1,irv2}, the coefficients should be reevaluated due to different material compositions on each terrestrial planet.

Despite its current limitations, this study presents a valuable tool for constraining theories of gravity. With a relatively simple Earth model, we have successfully constrained the most popular Ricci-based gravities and scalar-tensor theories up to $2\sigma$. Notably, our approach considers matter and its properties, avoiding assumptions like vacuum, dust, or simple equations of state commonly made in other approaches. By encoding these properties into seismic wave data, we can better control uncertainties related to matter description and modified gravity effects, leading to more stringent bounds on theory parameters.

Moreover, continuous efforts to refine and enhance the method, as discussed earlier, are expected to yield even more robust constraints. The research is currently advancing in this promising direction. Despite the challenges encountered, the constraints obtained in this study hold substantial potential to significantly narrow down the range of plausible alternative gravity theories.

\section*{Acknowledgements}
 AW acknowledges financial support from MICIN (Spain) {\it Ayuda Juan de la Cierva - incorporac\'ion} 2020 No. IJC2020-044751-I. \\
 AW would like to express her gratitude to CEICO, the Institute of Physics of the Czech Academy of Sciences, for their hospitality during the final stage of this work.\\

\end{document}